\documentclass{emulateapj}

\def \vv#1{\mathbf{#1}}

\def\pfe#1#2#3{\!\left(\!\frac{#1}{#2}\!\right)^{\!\!#3}}
\def \be {\begin{equation}}
\def \ee {\end{equation}}

\def\Deltae {\Delta e}
\def\Deltax {\Delta x}
\def\m {m}
\def\M {m_0}
\def\p {1}
\def\pp {{}}
\def\c {2}
\def\ij {i}

\def\gr {0}
\def\vpi {\varpi}

\def\cte {Cte}

\def\Kl {K'}

\def\crm{\cr\noalign{\medskip}}

\def \llabel#1{\label{#1}}

\def\apj{Astrophys.\ J. } 
\def\apjl{Astrophys.\ J. }

\def\aap{Astron.\ Astrophys. }

\def\gjras{Geophys.\ J.\ R.\ Astron.\ Soc. }

\def\mnras{Mon.\ Not.\ R.\ Astron.\ Soc. }

\def\nat{Nature }

\shorttitle{Pumping the eccentricity by tidal effect}
\shortauthors{A.C.M. Correia et al.}

\begin{document}


\title{Pumping the eccentricity of exoplanets by tidal effect} 


\author{Alexandre C.M. Correia}
\affil{Department of Physics, I3N, University of Aveiro, Campus de
Santiago, 3810-193 Aveiro, Portugal;}
\affil{Astronomie et Syst\`emes Dynamiques, IMCCE-CNRS UMR8028, 
77 Av. Denfert-Rochereau, 75014 Paris, France}
\email{correia@ua.pt} 

\author{Gwena\"el Bou\'e}
\affil{Centro de Astrof\'\i sica, Universidade do Porto, Rua das Estrelas,
	      4150-762 Porto, Portugal;}
\affil{Astronomie et Syst\`emes Dynamiques, IMCCE-CNRS UMR8028, 
77 Av. Denfert-Rochereau, 75014 Paris, France}

\and

\author{Jacques Laskar}
\affil{Astronomie et Syst\`emes Dynamiques, IMCCE-CNRS UMR8028, 
77 Av. Denfert-Rochereau, 75014 Paris, France}


\begin{abstract}
Planets close to their host stars are believed to undergo significant tidal interactions, leading to a progressive damping of the orbital eccentricity.
Here we show that, when the orbit of the planet is excited by an outer
companion, tidal effects combined with gravitational interactions may give rise to a secular increasing drift on the eccentricity.
As long as this secular drift counterbalances the damping effect, the eccentricity can increase to high values.
This mechanism may explain why some of the moderate close-in exoplanets are 
observed with substantial eccentricity values.
\end{abstract}





\section{Introduction}

Close-in exoplanets, 
as for Mercury, Venus and the majority of the natural satellites in the Solar
System, are supposed to undergo significant tidal interactions, resulting that
their spins and orbits are slowly modified. 
The ultimate stage for tidal evolution is the synchronization of the spin and
the circularization of the orbit \citep[e.g.][]{Correia_2009}.
The spin evolves in a shorter time-scale than the orbit, so long-term studies on the tidal evolution of exoplanets usually assume that their rotation is synchronously locked, and therefore
limit the evolution to the orbits.
However, these two kinds of evolution cannot be dissociated because the total
angular momentum must be conserved.
Synchronous rotation can only occur when the eccentricity is very close
to zero. Otherwise, the rotation rate tends to be locked with the orbital speed at
the periapsis, because tidal effects are stronger when the two bodies are closer
to each other.
In addition, in presence of a companion body, 
{the eccentricity undergoes oscillations \citep[e.g.][]{Mardling_2007}, and}
the rotation rate of the planet shows 
variations that follow the eccentricity \citep{Correia_Laskar_2004}.
As a consequence, some unexpected behaviors can be observed,
such as a secular increase of the eccentricity.
In this Letter we provide a simple averaged model for  the orbital and spin
evolution of an exoplanet with a companion (Sect.\,2),  
and apply it to the HD\,117618 planetary system (Sect.\,3). 
We then give an explanation for the eccentricity pumping (Sect.\,4),
and derive some conclusions (Sect.\,5).

\section{The model}

We consider here a system consisting of a central star of mass $\M$, an inner planet of mass $\m_\p$, and an outer companion of mass $\m_\c$.
We use Jacobi canonical coordinates, with $\vv{r_\p} $ being the position of
$\m_\p$ relative to $\M$, and $ \vv{r_\c} $ the position of $ \m_\c $
relative to the center of mass of $ \m_\p $ and $ \M $.
We further assume that $|\vv{r_\p}|  \ll  |\vv{r_\c}|$, 
that the system is coplanar, and that the obliquity of the planet is zero.
The inner planet is considered an oblate ellipsoid with gravity field coefficients given by $J_2$, rotating about the axis of maximal inertia, 
with rotation rate $\omega$, such that \citep[e.g.][]{Lambeck_1988}
\be
J_2 = k_2 \frac{\omega_\pp^2 R_\pp^3}{3 G \m_\p} \ .
\llabel{101220a}
\ee
$G$ is the gravitational constant, $R_\pp$ is the radius of the planet, and
$ k_2 $ is the second Love number for potential.

Since we are interested in the secular behavior of the system, we average the motion equations over the mean anomalies of both orbits.
The averaged potential, quadrupole-level for the spin \citep[e.g.][]{Correia_Laskar_2010},  octopole-level for the orbits \citep[e.g.][]{
Lee_Peale_2003,Laskar_Boue_2010}, and with general relativity corrections \citep[e.g.][]{Touma_etal_2009} is given by:

\begin{eqnarray}
U & =&  -  C_0 (1-e_\p^2)^{-1/2} - C_1 (1-e_\p^2)^{-3/2}   \crm & &
- C_2 \frac{(1 + \frac{3}{2} e_\p^2)}{(1-e_\c^2)^{3/2}}
+ C_3 \frac{e_\p e_\c (1 + \frac{3}{4} e_\p^2)}{(1-e_\c^2)^{5/2}} \cos \vpi 
\ , \llabel{090514a}
\end{eqnarray}
where 
\be
C_0 = \frac{3 \beta_\p G^2 (\M + \m_\p)^2}{a_\p^2 c^2} \ , \quad
C_1 = \frac{G \M  \m_\p J_2 R_\pp^2}{2 a_\p^3}  \llabel{110816a} \ ,
\ee
\be
C_2 = \frac{G \beta_\p \m_\c  a_\p^2}{4 a_\c^3 }  \ , \quad
C_3 = \frac{15 G \beta_\p \m_\c a_\p^3}{16 a_\c^4} \frac{(\M-\m_\p)}{\M + \m_\p} \llabel{110816c} \ .
\ee
$a_\ij$ is the semi-major axis (that can also be expressed using the mean motion $n_\ij $),
$e_\ij$ is the eccentricity, and $ \vpi = \vpi_\p - \vpi_\c $ is the difference between the longitudes of the periastron, $ \vpi_\ij $.
We also have $ \beta_\p = \M m_\p / (\M + \m_\p) $, and  $ \beta_\c = (\M + \m_\p) \m_\c / (\M + \m_\p + \m_\c) $. 

The contributions to the orbits are easily obtained using the Lagrange planetary equations \citep[e.g.][]{Murray_Dermott_1999}:
\be
\dot e_\ij = \frac{\sqrt{1-e_\ij^2}}{\beta_\ij n_\ij a_\ij^2 e_\ij} \frac{\partial U}{\partial \vpi_\ij} \ , \quad
\dot \vpi_\ij = - \frac{\sqrt{1-e_\ij^2}}{\beta_\ij n_\ij a_\ij^2 e_\ij} \frac{\partial U}{\partial e_\ij} \llabel{110816d} \ .
\ee
Thus,
\be
\dot e_\p = - \nu_{31}  \frac{e_\c (1 + 3/4 e_\p^2) \sqrt{1-e_\p^2}}{(1-e_\c^2)^{5/2}} \sin \vpi \llabel{110816h} \ ,
\ee
\be
\dot e_\c = \nu_{32} \frac{e_\p (1 + 3/4 e_\p^2)}{(1-e_\c^2)^{2}} \sin \vpi \llabel{110816i} \ ,
\ee
and
\begin{eqnarray}
\dot \vpi &=& \frac{\nu_\gr}{ (1-e_\p^2)} + \frac{\nu_1 \, x_\pp^2}{(1-e_\p^2)^2} \crm 
&+& \nu_{21} \frac{\sqrt{1-e_\p^2}}{ (1-e_\c^2)^{3/2}} - \nu_{22} \frac{ (1+\frac{3}{2} e_\p^2)}{ (1-e_\c^2)^2}  \crm 
&-& \nu_{31} \frac{e_\c \sqrt{1-e_\p^2} (1+\frac{9}{4} e_\p^2)}{e_\p (1-e_\c^2)^{5/2}} \cos \vpi \crm 
&+& \nu_{32} \frac{e_\p (1+\frac{3}{4} e_\p^2) (1+4 e_\c^2)}{e_\c (1-e_\c^2)^3} \cos \vpi
\llabel{110819a} \ ,
\end{eqnarray}
where $ x_\pp = \omega_\pp / n_\p $, and the constant frequencies
\be
\nu_\gr = 3 n_\p \pfe{n_\p a_\p}{c}{2} \ , \llabel{110817f}
\ee
\be
\nu_1 = n_\p \frac{k_2}{2} \frac{\M + \m_\p}{\m_\p} \pfe{R_\pp}{a_\p}{5} \ , \llabel{110817a}
\ee
\be
\nu_{21} =  n_\p \frac{3}{4} \frac{\m_\c}{\M + \m_\p} \pfe{a_\p}{a_\c}{3} \ , \llabel{110817b}
\ee
\be
\nu_{22} =  n_\c \frac{3}{4} \frac{\M \m_\p}{(\M +\m_\p)^2} \pfe{a_\p}{a_\c}{2}  \ , \llabel{110817c}
\ee
\be
\nu_{31} =  n_\p \frac{15}{16} \frac{\m_\c}{\M+\m_\p} \frac{\M-\m_\p}{\M+\m_\p} \pfe{a_\p}{a_\c}{4} \ , \llabel{110817d}
\ee
\be
\nu_{32} =  n_\c \frac{15}{16} \frac{\M \m_\p}{(\M+\m_\p)^2} \frac{\M-\m_\p}{\M+\m_\p} \pfe{a_\p}{a_\c}{3} \ . \llabel{110817e}
\ee
Notice that the variations in $e_\p$ and $e_\c$ (Eqs.\,\ref{110816h},
\ref{110816i}) are related by the conservation of the total angular momentum
(after dividing by $ m_\p n_\c a_\c^2 $):
\be
\xi\frac{\omega_\pp}{n_\c} \pfe{R_\pp}{a_\c}{2} + \frac{\beta_\p}{\m_\p} \frac{n_\p}{n_\c} \pfe{a_\p}{a_\c}{2} \!\!\sqrt{1-e_\p^2} + \frac{\beta_\c}{\m_\p} \!\sqrt{1-e_\c^2} = \cte \ , \llabel{110819b}
\ee
where $ \xi $ is a structure coefficient.
The conservative system (Eq.\,\ref{090514a}) can thus be reduced to one degree of freedom.


In our model, we additionally consider tidal dissipation raised by the central star on the inner planet.
The dissipation of the mechanical energy of tides in the planet's interior is responsible for a time delay $\Delta t_\pp$ between the initial perturbation and the maximal deformation. As the rheology of planets is badly known, the exact dependence of $\Delta t_\pp$ on the tidal frequency is unknown.
Several models exist \citep[for a review see][]{Correia_etal_2003,Efroimsky_Williams_2009}, but for simplicity we adopt here a model with constant $\Delta t_\pp$, 
which can be made linear \citep{Singer_1968,Mignard_1979}.
The contributions to the equations of motion are given by \citep[e.g.][]{Correia_2009}:
\be
\frac{\dot \omega_\pp}{n_\p} = - K_\pp
\left( f_1(e_\p) x_\pp - f_2(e_\p)  \right) \ , \llabel{090515a}
\ee
\be
\frac{\dot a_\p}{a_\p} = 2 \Kl_\pp \,
\left( f_2(e_\p) x - f_3(e_\p) \right) \ , \llabel{090515b}
\ee
\be\dot e_\p = 9 \Kl_\pp \left( \frac{11}{18} f_4(e_\p) x - f_5(e_\p) \right) e_\p \ ,
\llabel{090515c}
\ee
where
\be
K_\pp =  n_\p \frac{3 k_2}{\xi Q} \frac{\M \beta_\p}{\m_\p^2} \pfe{R_\pp}{a_\p}{3} \ , \quad \Kl_\pp = \frac{K_\pp}{1/\xi}
\frac{\m_\p}{\beta_\p} \pfe{R_\pp}{a_\p}{2} \llabel{090514m} \ , 
\ee
$ Q_\pp^{-1} = n_\p \Delta t_\pp $,
and
$ f_1(e) = (1 + 3e^2 + 3e^4/8) / (1-e^2)^{9/2} $, 
$ f_2(e) = (1 + 15e^2/2 + 45e^4/8 + 5e^6/16) / (1-e^2)^{6} $,
$ f_3(e) = (1 + 31e^2/2 + 255e^4/8 + 185e^6/16 + 25e^8/64) / (1-e^2)^{15/2} $,
$ f_4(e) = (1 + 3e^2/2 + e^4/8) / (1-e^2)^5 $,
$ f_5(e) = (1 + 15e^2/4 + 15e^4/8 + 5e^6/64) / (1-e^2)^{13/2} $.

We neglect the effect of tides over the longitude of the periastron, as well as the flatenning of the central star. Their effect is only to add a small supplementary frequency to $\vpi_\p$, similar to the contributions from the general relativity 
\citep[for a complete model see][]{Correia_etal_2011}.

Under the effect of tides alone,
the equilibrium rotation rate, obtained when $ \dot \omega_\pp = 0 $, is attained for (Eq.\,\ref{090515a}):
\be
\frac{\omega_\pp}{n_\p} = f(e_\p) = \frac{f_2(e_\p)}{f_1(e_\p)} = 1 + 6 e_\p^2 + {\cal O}(e_\p^4)
 \ . \llabel{090520a}
\ee
Usually $ \Kl_\pp \ll K_\pp $, so tidal effects modify the rotation rate much faster than the orbit. It is thus tempting to replace the equilibrium rotation in expressions (\ref{090515b}) and (\ref{090515c}).
With this simplification, one obtains always negative contributions for $ \dot a_\p $ and $ \dot e_\p $ \citep{Correia_2009}, 
\be
\frac{\dot a_\p}{a_\p} = - 7 K_\pp' \, f_6(e_\p) e_\p^2
\ , \llabel{090522a}
\ee
\be
\dot e_\p = - \frac{7}{2} K_\pp' f_6(e_\p) (1-e_\p^2) e_\p 
\ , \llabel{090522b}
\ee
with 
$ 
f_6 (e) = (1 + 45e^2/14 + 8e^4 + 685e^6/224 + 255e^8/448 + 25e^{10}/1792)
(1-e^2)^{-15/2} / (1 + 3e^2 + 3e^4/8) . \llabel{090527a}
$
Thus, the semi-major axis and the eccentricity can only decrease until the orbit of the planet becomes circular (Fig.\,\ref{fig1}a).
However, planet-planet interactions can produce eccentricity
oscillations with a period shorter, or comparable to the damping timescale of the
spin. In that case, the expression (\ref{090520a}) is not satisfied
and multi-planetary systems may show non-intuitive eccentricity evolutions,
such as eccentricity pumping of the inner orbit ($e_\p $ increases while
$e_\c$ decreases).

\section{Application to exoplanets}

As an illustration of the eccentricity pumping, we apply our model to the HD\,117618 system.
This Sun-like star ($\M \approx M_\odot$) has been reported to host a single Saturn-like planet on a eccentric orbit \citep{Butler_etal_2006}.
The residuals of the best fitted solution to the observational data are 5.5\,m/s, so we assume that any additional companion with a doppler shift semi-amplitude smaller than this value is presently undetected, that is, any planet with $ \m_\c < 0.2 \; M_{J} $ and $ a_\c > 1.4 $\,AU.

In our simulations we adopt for the observed planet {the same geophysical parameters as for Saturn,} $ R_\pp =  R_{Sat} $, $ k_2 = 1/2 $, $ \xi_\pp = 1/5 $, and a dissipation time lag $ \Delta t_\pp = 200 $\,s ({which is equivalent to $ Q_\pp \approx 3 \times 10^4$}).
Since the semi-major axis of the planet undergoes tidal dissipation, its value was certainly larger when the system formed.
We then adopt $ a_\p = 0.25 $\,AU as initial value for all simulations.
The initial eccentricity is chosen such that $ e_\p \approx 0.4 $ when $ a_\p = 0.175 $, the present observed values (Tab.\,\ref{table1}).
We further assume initial $ \vpi = 180^\circ $, and $ 2 \pi / \omega_\pp = 50 $\,day, {that quickly evolves near the equilibrium rotation (Eq.\,\ref{090520a}).} 

\begin{table}
\begin{center}
\caption{Single planetary systems with $0.1 < a_\p < 0.3 $ and $e_\p > 0.3$. \llabel{table1}}
\begin{tabular}{lcccccc}
\tableline
\tableline
Star & $a_\p$ & $e_\p$ & $\m_\p$ & $\M$ & Age & $\tau $  \\
(name) & (AU) &   & ($M_{J}$) & ($M_\odot$) & (Gyr) & (Gyr)  \\
\tableline
HD\,108147 & 0.102 & 0.53 & 0.26 & 1.19 & 2.0 & 0.01 \\
CoRoT-10   & 0.105 & 0.53 & 2.75 & 0.89 & 3.0 & 0.24 \\
HD\,33283  & 0.145 & 0.48 & 0.33 & 1.24 & 3.2 & 0.34 \\
HD\,17156  & 0.163 & 0.68 & 3.19 & 1.28 & 3.4 & 0.44 \\
HIP\,57050 & 0.164 & 0.31 & 0.30 & 0.34 & $-$ & 39.4 \\
HD\,117618 & 0.176 & 0.42 & 0.18 & 1.05 & 3.9 & 2.06 \\
HD\,45652  & 0.228 & 0.38 & 0.47 & 0.83 & $-$ & 93.3 \\
HD\,90156  & 0.250 & 0.31 & 0.06 & 0.84 & 4.4 & 35.8 \\
HD\,37605  & 0.260 & 0.74 & 2.84 & 0.80 & 10.7& 10.6 \\
HD\,3651   & 0.284 & 0.63 & 0.20 & 0.79 & 5.1 & 15.5 \\
\tableline
\end{tabular}
\tablecomments{Data is taken from http://exoplanet.eu/  \\  $ \tau = \frac{2 \m_\p a_\p^8 (1-e_\p^2)^8}{ \Delta t_\pp 21 k_2 G \M^2 R_\pp^5}$, with $ \Delta t_\pp = 10^2 $\,s \citep{Correia_Laskar_2010B}. }
\end{center}
\end{table}

\begin{figure*}
\includegraphics[width=17cm]{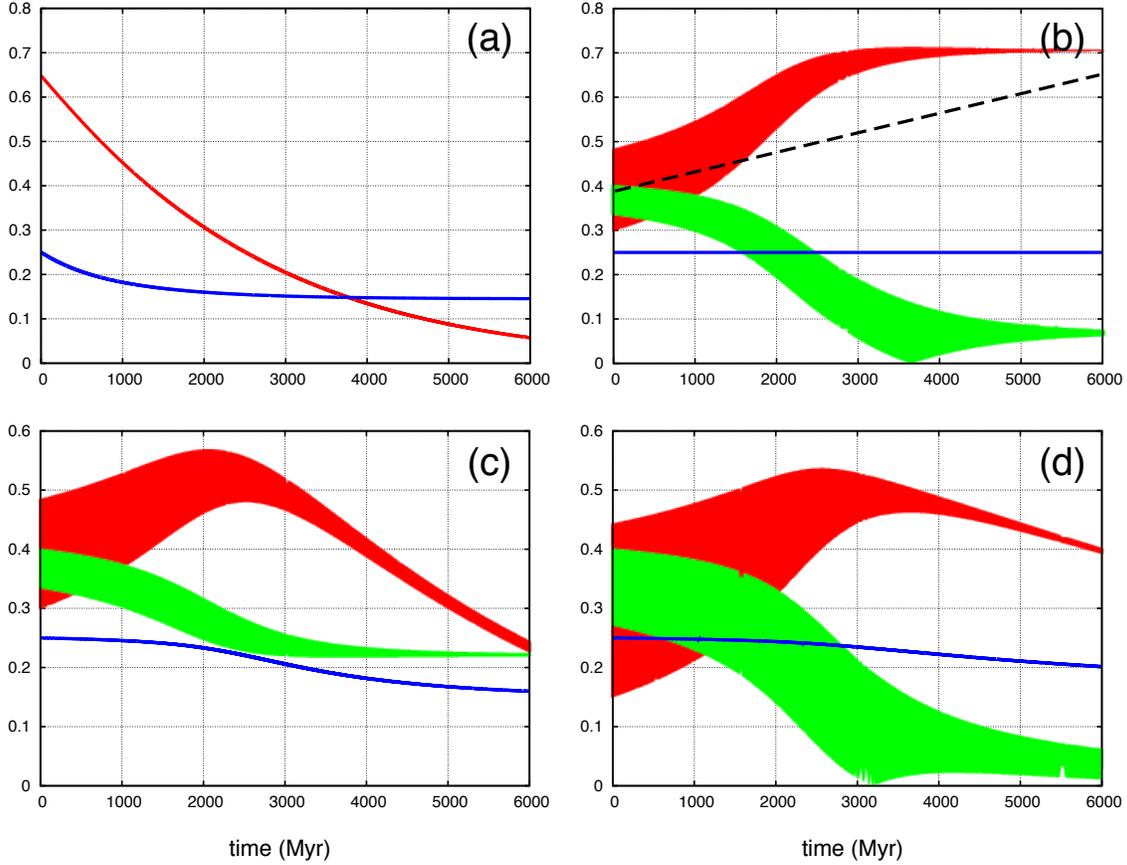}
\caption{Long-term evolution of the HD\,117618 system in different situations. We show the semi-major axis $a_\p$ in AU (blue), and the eccentricities $e_\p$ (red) and $e_\c$ (green). {\bf (a)} without a companion; {\bf (b)} with a $ \m_\c = 0.2\,M_{J} $ companion at  $ a_\c = 1.8 $\,AU, but without dissipation on the orbit; {\bf (c)} same as (b), but with a full model; {\bf (d)} with a $ \m_\c = 0.15\,M_{J} $ companion at  $ a_\c = 1.4 $\,AU, with a full model.
The dashed line gives the slope of the linear approximation (Eq.\,\ref{110902b}).
 \llabel{fig1}}
\end{figure*}

In absence of a companion, the eccentricity and the semi-major axis are damped following an exponential decay \citep{Correia_Laskar_2010B}, and the present configuration is attained after 1\,Gyr (Fig.\,\ref{fig1}a).
At present the observed eccentricity would be around 0.1, and we still needed to
explain the high initial value near 0.7.

We now add a companion to the system with $ \m_\c = 0.2\,M_{J} $, $ a_\c = 1.8 $\,AU, and $ e_\c = 0.4 $, 
and set $ e_\p = 0.3 $.
At first, we only consider dissipation in the spin (Eq.\ref{090515a}) and neglect its effect on the orbit (Eqs.\,\ref{090515b},\,\ref{090515c}), in order to highlight the eccentricity pumping (Fig.\,\ref{fig1}b).
We then clearly observe this effect, the eccentricity of the inner planet rising up to 0.7.
We also observe that the eccentricity of the outer planet is simultaneously damped, because of the conservation of the total angular momentum (Eq.\,\ref{110819b}).

Orbital and spin evolution cannot be dissociated, so we then integrate the full system (Fig.\,\ref{fig1}c).
We observe that the initial behavior of the system is identical to the situation without dissipation on the orbit (Fig.\,\ref{fig1}b).
However, as the eccentricity increases, the inner planet comes closer to the star at periastron, and tidal effects on the orbit become stronger.
As a consequence, the semi-major axis decreases and the damping effect on the eccentricity (Eq.\,\ref{090515c}) overrides the pumping drift. 
The system ultimately evolves into a circular orbit.
The present configuration is attained around 4\,Gyr of evolution, which is compatible with the present estimated age of the star.
The pumping effect is then responsible for a delay in the final evolution of planetary systems and may explain the high values observed for some of them (Tab.\,\ref{table1}).

Finally, we repeat the integration of the full system, but  with a smaller-mass companion $ \m_\c = 0.15\,M_{J} $ at $ a_\c = 1.4 $\,AU. 
The companion eccentricity is still $ e_\c = 0.4 $, but the inner planet now begins with $ e_\p = 0.15 $ (Fig.\,\ref{fig1}d).
The initial evolution is still similar to the previous simulation (Fig.\,\ref{fig1}c), except that the eccentricity oscillations of both planets are higher, because the orbits are closer.
Since the companion mass is smaller, its eccentricity also decreases more than
before, and reaches zero around 3\,Gyr.
At this stage, the angle $\vpi$ stops circulating, and begins librating around $180^\circ$.
The two orbits are then tightly coupled and evolve together, showing an identical behavior to close-in planets with a few days of orbital period \citep[e.g.][]{Mardling_2007,
Laskar_etal_2011}.
As a consequence, the evolution time-scale is much longer, allowing the inner planet to maintain
high eccentricity for longer periods of time (Fig.\,\ref{fig1}d). 
{The present eccentricity is only observed after 6\,Gyr.} 

\vskip1truecm

\section{Eccentricity pumping}

In order to understand the unexpected behavior of the eccentricity during the initial stages of the
evolution, 
we can perform some simplifications in the equations of motion
without loss of generality (Sect.\,2).
We can neglect tidal effects on orbital quantities  (Eqs.\,\ref{090515b},\,\ref{090515c}), which is justified since $ \Kl_\pp\ll K_\pp $ (Eq.\ref{090514m}).
The only contribution of tides is then on the rotation rate (Eq.\ref{090515a}).
The semi-major axis and the mean motion are thus constant, and the eccentricity only varies due to the gravitational perturbations (Eq.\,\ref{110816h}).
In addition, we linearize the set of equations of motion in the vicinity
of the averaged values of $x$, $e_\p$, and $e_\c$. Let $x=x_0+\delta x$,
where $x_0$ is the solution of (\ref{090520a}), $e_\p = e_{\p 0} +
\delta e_\p$, and $e_\c = e_{\c 0} + \delta e_\c$. In the following,
$\delta e_\c$ is expressed as a function of $\delta x$ and $\delta e_\p$
using the conservation of the angular momentum (Eq.\ref{110819b}).
Then, the set of equations of motion (\ref{110816h}, \ref{110819a}, \ref{090515a}) reduces to:
\be
\delta \dot e_\p = - A \sin \vpi \ , \llabel{110812b}
\ee
\be
\dot \vpi = g + g_x \delta x + g_e \delta e_\p\ , \llabel{110812c}
\ee
\be
\delta \dot x = -\nu_x \delta x + \nu_e \delta e_\p \ , \llabel{110812d}
\ee
with 
\be 
A = \nu_{31} \frac{e_{\c 0} (1 + 3/4 e_{\p 0}^2) \sqrt{1-e_{\p 0}^2}}{(1-e_{\c
0}^2)^{5/2}} 
\ , 
\ee
\begin{eqnarray} 
g &=& \frac{\nu_0}{(1-e_{\p 0}^2)} + \frac{\nu_1 x_0^2}{(1-e_{\p 0}^2)^2} 
\crm && 
+ \nu_{21} \frac{\sqrt{1-e_{\p 0}^2}}{(1-e_{\c 0}^2)^{3/2}} - \nu_{22}
\frac{(1+3 e_{\p 0}^2/2)}{(1-e_{\c 0}^2)^2} \ ,
\end{eqnarray}
\be g_x = \nu_1 \frac{2 x_0}{(1-e_{\p 0}^2)^2} 
\ ,
\ee
\begin{eqnarray}
g_e &=& \nu_0 \frac{2 e_{\p 0}}{(1-e_{\p 0}^2)^2} + \nu_1 \frac{4 x_0^2 e_{\p
0}}{(1-e_{\p 0}^2)^3} 
- \nu_{22} \frac{3 e_{\p 0}}{(1-e_{\c 0}^2)^2}
\crm && - \nu_{21} \frac{e_{\p 0}}{\sqrt{1-e_{\p 0}^2} (1-e_{\c 0}^2)^{3/2}} 
\ , 
\end{eqnarray}
\be 
\nu_x =  K_\pp f_1(e_{\p 0}) \ ,
\ee
\be 
\nu_e = -K_\pp ( f'_1 (e_{\p 0}) x_0 - f'_2(e_{\p 0}) ) \ ,
\ee
where
$f'_1(e) = 15 (e+3 e^3/2 +e^5/8) / (1-e^2)^{11/2} $, and
$f'_2(e) =  3 (9e+65e^3/2+125e^5/8+5e^7/8) / (1-e^2)^{7} $.
We neglected the octupole terms since $ \nu_{3\ij} \ll \nu_{2\ij} $, the
contributions from $\delta e_\c $, and assumed that $e_{\ij 0} \ne 0$.

At first order, the precession of the periastron is constant
$ \dot \vpi \simeq g $, 
and the eccentricity is simply given from expression (\ref{110812b}) as
\be
\delta e_\p = \Deltae \cos  (g t + \vpi_0) \ , \llabel{110812e}
\ee
where $ \Deltae = A / g $. 
That is, the eccentricity $e_\p$ presents periodic variations around an equilibrium value $ e_{ \p 0} $, with amplitude $ \Deltae $ and frequency $ g $.
Since $ g_x \delta x, g_e \delta e_\p \ll g$, the above solution for the eccentricity can be adopted as the zeroth order solution of the system of equations (\ref{110812b}$-$\ref{110812d}).
With this approximation, the equation of motion of $\delta x$ (\ref{110812d}) becomes that of a driven harmonic oscillator whose the steady state solution is
\be
\delta x = \Deltax \cos ( g t + \vpi_0 - \phi) \ , \llabel{110920a}
\ee
with $ \Deltax = \nu_e \Deltae / \sqrt{\nu_x^2 + g^2 } $, 
and $\sin \phi = g/\sqrt{\nu_x^2 + g^2} $. The rotation rate 
thus presents an oscillation identical to the eccentricity (Eq.\ref{110812e}), but with smaller amplitude and delayed by an angle $\phi$
\citep[see][]{Correia_2011}.
Using the above expression in equation (\ref{110812c}) and integrating, gives for the periastron:
\be
\vpi = g t + \vpi_0 + \frac{g_x}{g} \Deltax \sin (g t + \vpi_0 - \phi)
+ \frac{g_e}{g} \Deltae \sin (g t + \vpi_0)
\llabel{110812g} \ .
\ee
Finally, substituting in expression (\ref{110812b}) and using the approximation $ g_x \Deltax, g_e \Deltae \ll g$ gives
\begin{eqnarray}
\delta \dot e_\p 
& \approx & - A \sin (g t + \vpi_0)
- \frac{g_e A}{g} \Deltae \sin (g t + \vpi_0) \cos  (g t + \vpi_0) \crm 
& & - \frac{g_x A}{g} \Deltax \sin (g t + \vpi_0 - \phi) \cos  (g t + \vpi_0) 
  \llabel{110812h} \ ,
\end{eqnarray}
or, combining the two products of periodic functions,
\begin{eqnarray}
\delta \dot e_\p  & = & - A \sin (g t + \vpi_0) 
 - \frac{g_x A}{2 g} \Deltax \sin (2 g t + 2 \vpi_0 - \phi) \crm
& & - \frac{g_e A}{2 g} \Deltae \sin (2 g t + 2 \vpi_0)  +  \frac{g_x A}{2 g} \Deltax \sin \phi
\llabel{110812i} \ .
\end{eqnarray}
The two middle terms in the above equation can be neglected since they are
periodic and have a very small amplitude ($ g_x \Deltax, g_e \Deltae \ll g $). However, the last term in $\sin \phi$ is constant and it adds a small drift to the eccentricity,
\be
< e_\p > = \frac{\nu_e g_x A^2}{2 g (\nu_x^2 + g^2 )}  t \ . \llabel{110902b}
\ee
The drift is maximized for $ g \sim \nu_x $,  it vanishes for weak dissipation ($ \Deltax \rightarrow 0 $), but also for strong dissipation ($\phi \rightarrow 0$).
Note that the phase lag $\phi$ between the eccentricity (Eq.\,\ref{110812e}) and the rotation variations (Eq.\,\ref{110920a}) is essential to get a drift on the eccentricity.
That is why the eccentricity pumping was never observed in previous studies that did not take into account the spin evolution.

The major difference when we consider the full non-linearized problem
is that the drift (Eq.\,\ref{110902b}) cannot grow indefinitely.
Indeed, when the eccentricity reaches high values, the drift vanishes (Fig.\,\ref{fig1}b).
Moreover, tidal effects are also enhanced for high eccentricities and counterbalance the drift (Eq.\,\ref{090515c}).
As a consequence, the drift on the eccentricity is never permanent, although it can last for the age of the system (Fig.\,\ref{fig1}c,d).

In order to observe the pumping effect, 
the eccentricity should not be damped, while 
the damping timescale of the spin of the planet should be of the order
of the period of eccentricity oscillation.
This is valid for gaseous planets roughly within $0.1 < a_\p < 0.3 $.
About half of the planets in this range present eccentricities higher than 0.3 (Tab.\,\ref{table1}).
This is somehow unexpected, since the time-scale for damping the eccentricity ($ \tau \sim 1/\Kl$) is shorter than
the age of those systems (Tab.\,\ref{table1}, Fig.\,\ref{fig1}a).
Thus, unless the initial eccentricity of those planets was extremely high, one may suspect of the existence of undetected companions that help the eccentricity to maintain the present values.

\section{Conclusion}


In this Letter we have shown that, under some particular initial conditions,
orbital and spin evolution cannot be dissociated.
Indeed, counterintuitive behaviors can be observed, such as the secular augmentation of the eccentricity.
This effect can last over long time-scales and may explain the high eccentricities observed for moderate close-in planets. 

The variations in the flattening of the inner planet due to rotation (Eq.\,\ref{101220a}) is a key element to pump the eccentricity by means of $g_x$ (Eq.\,\ref{110902b}).
We have considered an instantaneous response to the rotation in $J_2$, but a time delay until the maximum deformation is reached is to be expected.
This may increase the phase lag between the eccentricity forcing (Eq.\,\ref{110812e}) and the precession angle oscillations (Eq.\,\ref{110812g}).
It results that the drift effect on the eccentricity can be even more pronounced than the one presented here.

\acknowledgments

We acknowledge support from the PNP-CNRS, 
the FCT (grant PTDC/CTE-AST/098528/2008),
and
the European Research Council.



\end{document}